\documentstyle[epsf]{l-aa}

\newcommand{\aaa}{A\&A}      % Astronomy and Astrophysics
\newcommand{\apj}{ApJ}       % Astrophysical Journal
\newcommand{\bm}[1]{\mbox{{\boldmath $#1$}}}
\newcommand{\non}{\nonumber}
\newcommand{\fr}[2]{\frac{\displaystyle #1}{\displaystyle #2}}

\newcommand{\pder}[3]{\fr{{\partial}^{#3} {#1}}{{\partial} {#2}^{#3}}}
\newcommand{\cder}[3]{\fr{D^{#3} {#1}}{D {#2}^{#3}}}
\newcommand{\const}{\mbox{ const\ $\;$}}

\newcommand{\paral}{\parallel}
\newcommand{\cross}{\times}
\renewcommand{\b}{\!\!\!}
\newcommand{\km}{{\rm \ km}}
\newcommand{\cm}{{\rm \ cm}}
\newcommand{\pc}{{\rm \ pc}}

\newcommand{\s}{{\rm \ s}}

\newcommand{\Myr}{{\rm \ Myr}}

\renewcommand{\Im}{\rm Im}
\renewcommand{\Lambda}{H}

\begin{document}

\thesaurus{02(02.13.1; 02.09.1; 11.13.2; 11.19.2; 11.09.4; 09.11.1)}

\title{The galactic dynamo effect due to Parker-shearing instability of
magnetic flux tubes}
\subtitle{I. General formalism and the linear approximation.}

\author{M. Hanasz \inst{1} \and H. Lesch \inst{2}}

\institute{Institute of Astronomy, Nicolaus Copernicus University,
 ul. Chopina 12/18, PL-87-100 Torun, Poland.
\and
University Observatory, M\"unchen University,
Scheinerstr. 1, 81679 M\"unchen , Germany}

\offprints{M. Hanasz, e-mail: {\em mhanasz@astri.uni.torun.pl} }

\date{Received July 31, 1996/ accepted October 30, 1996}

\maketitle

\markboth{M.  Hanasz et al.: The galactic dynamo effect due to Parker-shearing
instability of magnetic flux tubes }{}

\begin{abstract}

In this paper we investigate the idea of Hanasz \& Lesch 1993 that the galactic
dynamo effect is due to the Parker instability of magnetic flux tubes.  In
addition to the former approach, we take into account more general physical
conditions in this paper, by incorporating cosmic rays and differential forces
due to the axisymmetric differential rotation and the density waves as well.

We present the theory of slender magnetic flux tube dynamics in the thin flux
tube approximation and the Lagrange description.  This is the application of
the formalism obtained for solar magnetic flux tubes by Spruit (1981), to the
galactic conditions.  We perform a linear stability analysis for the
Parker-shearing instability of magnetic flux tubes in galactic discs and then
calculate the dynamo coefficients.

We present a number of new effects which
are very essential for cosmological and contemporary evolution of galactic
magnetic fields.  First of all we demonstrate that a very strong dynamo
$\alpha$-effect is possible in the limit of weak magnetic fields in presence of
cosmic rays.  Second, we show that the differential force resulting from
axisymmetric differential rotation and the linear density waves causes that the
$\alpha$-effect is essentially magnified in galactic arms and switched off in
the interarm regions.  Moreover, we predict a non-uniform magnetic field in
spiral arms and well aligned one in interarm regions.  These properties are
well confirmed by recent observational results by Beck \& Hoernes (1996)


\end{abstract}

\section{Introduction}

The origin of galactic magnetic fields is a long standing problem in
theoretical astrophysics.  Virtually all spiral galaxies have magnetic fields
of a few $\mu$G which are spatially coherent over several parsecs (see Kronberg
1994; Beck et al.  1996 for recent reviews).  This implies that the average
magnetic energy density is comparable to the average energy densities of cosmic
rays and the interstellar medium, which in turn suggests that galactic magnetic
fields reached a state of saturation some unknown time in the past.  In face of
the Faraday rotation measurements of Wolfe et al.  (1992), which indicate
$\mu$G fields in damped Lyman$\alpha$ clouds at redshifts of 2 (the universe
was about 1-2 Gigayears old), the saturation has to be very rapid.  Since
galaxies are so large (and damped Lyman$\alpha$ clouds are supposed to be
progenitors of disc galaxies), and their constituent gases are highly
conducting, the dissipative time scale for these fields is considerably longer
than a Hubble time.  Either the magnetic fields are simply the result of some
pre-existing cosmological field (Kulsrud 1990) or they are spectacular examples
of fast dynamos, in which rapid magnetic reconnection takes over the role of
the traditional eddy diffusivity (Parker 1992)

Nowadays the regular magnetic fields in the discs of spiral galaxies are
usually considered to be the result of large-scale dynamo action, involving a
collective induction effect of turbulence ($\alpha$-effect) and differential
rotation.  Both mechanisms have been described in detail in a huge number of
articles (e.g.  Kronberg 1994; Beck et al.  1996 and references therein).  A
small seed field is exponentially amplified by the interplay of turbulent
motions perpendicular to the disc generating a radial magnetic field component
$B_{\rm r}$ from an azimuthal component $B_{\varphi}$.  Differential rotation
in the galactic disc then closes the loop by generating $B_{\rm \varphi}$ from
$B_{\rm r}$.  The magnetic field grows exponentially on time scales of the
order of a Gigayear (Camenzind and Lesch 1994).  The growth time raises the
problem of magnetic field amplification at high redshifts.  The typical seed
field strengths is of the order of $10^{-18}$ G and the typical present
magnetic field strength is of the order of $10^{-6}$ G.  The
e-folding times of the dynamo magnification process have to be of the order of a few
hundreds of Myr or even a few tens of Myr for the total time of growth 10 Gyr
or 1 Gyr respectively.  Such a short e-folding time seems to be unavailable
within the classical dynamo models (Lesch and Chiba 1995).

Whereas the differential rotation of galactic disc is a clearly observed
feature of spiral galaxies, there is no general agreement on the source or
value of $\alpha$.  The helicity $\alpha$ is connected with the interstellar
turbulence, which may be fed by several sources (warm gas, hot bubbles, cold
filaments, supernova explosions, cloud-cloud collisions, OB-associations, the
galactic fountain, etc...).  These violent motions agitate the interstellar
gases and drive turbulence.  The classical $\alpha$-effect measures the
field-aligned electromotive force resulting from magnetic field lines twisted
by the turbulence.  Originally the dynamics of these field lines is governed by
external turbulent motions like supernova explosions (Ruzmaikin, Sokoloff and
Shukurov 1988; Ferriere 1993).  However, attempts to compute the several components of the
helicity tensor $\alpha_{\rm ij}$ resulting from exploding supernovas or
superbubbles resulted in values considerably to small to allow for fast and
efficient dynamo action (Kaisig et al. 1993). { Nevertheless, the most
recent calculation   of the superbubble contribution to the helicity tensor by
Ferriere (1996) makes the superbubble model more  attractive.}
Beside that time scale problem, the recent observations by Beck and Hoernes
(1996) raise the question what kind of global process in a galactic disc can
explain the "magnetic arms" in NGC 6946.  They report the detection of two
highly polarized magnetic features which are 0.5 - 1 kpc wide and about 12 kpc
long and have greater symmetry than the optical arms.  These features cannot be
explained by standard dynamo models.

The usual dynamo models treat the magnetic field as a superposition of a
homogeneous mean field and a fluctuating, turbulent field component, which is
responsible for the turbulent diffusion and the helicity, as well.  We
considered in a first paper (Hanasz and Lesch 1993, hereafter HL'93)
 the Parker instability of magnetic flux tubes (Parker 1955, 1966, 1967a,b)  as the
source for the helicity, taking into account the observed complicated polarized
filaments, which resemble the complex structure of the interstellar gas
velocities and densities.  The concept of flux-tube dynamo is well known for
stars (Sch\"ussler 1993), but has never been developed for galaxies.

The Parker instability occurs in a composition of gas, magnetic field and
cosmic rays placed in a gravitational field, if the magnetic field and the
cosmic rays contribute to a pressure equilibrium of the system.  The given
pressure equilibrium of the flux tube implies that the density of the
composition of gas, magnetic field and cosmic radiation is smaller than the
density of the surrounding gas, which in the presence of a gravitational field
leads to buoyant field lines.  This situation is clearly realized in galactic
halos.  If the Parker instability operates, the gas together with the frozen-in
magnetic field and cosmic radiation captured by the magnetic field start to
ascend, forming bumps on the initially straight horizontal field lines.  The
portion of gas in the ascending area  starts
to sink along the field lines, forming the condensations of the gas in the
valleys and rarefaction in summits.

Taking into account the dragging effect of turbulence in the surrounding medium
we could show that the Parker instability provides a helicity of about
$\alpha\sim 0.6\, \km \s^{-1}$ and that the turbulent diffusion coefficient is
reduced by the effect of aerodynamic drag force, which causes an increase in
dynamo efficiency.
This means that a dynamo driven by Parker instability saturates
faster than a classical dynamo.  It also means that the magnetic field in a
galaxy will always stay in the saturated state, regulated by magnetic flux loss
via the Parker instability.  Similar ideas for a fast dynamo were proposed by
Parker (1992), who suggests a mechanism with rising and inflating magnetic
lobes driven by cosmic rays.

In our first article we did not consider the effects of the cosmic rays and the
underlying axisymmetric differentially rotating disc, including
non-axisymmetric density waves.  It is the aim of this contribution to show the
results of an extended version of a flux tube dynamo, including cosmic rays in
a differentially rotating disc with density waves.

The influence of the cosmic rays is obvious in face of galaxy evolution.
Simply the fact that typically 90\% of the gas in a galaxy has been transformed
into stars means that star formation was very efficient in the early phases of
evolution.  For the magnetic field evolution in the flux tube model this means
that the Parker instability was effectively driven by the large cosmic ray
flux, since a strong star formation is always accompanied by many supernova
explosions, i.e.  by efficient particle acceleration.  Since the cosmic ray
pressure can be assumed to be dominant in star forming galaxies (see also M82
(Reuter et al.  1992)), it regulates the efficiency of the Parker
instability.

The role of differentially rotating disc on the Parker instability is rather
complicated.  The differential forces introduce a new level of complexity to
the problem.  According to Foglizzo \& Tagger (1994, 1995, hereafter FT'94 and
FT'95) the action of differential rotation depends on the magnitude of shear
but also on the radial and vertical wavenumbers of the perturbation.  The
differential rotation can transiently stabilize mode of long vertical
wavelength, while the waves with short vertical wavelength are subject to both
the Parker and the transient shearing instabilities.  In the case of general 3D
distribution of magnetic field the linear coupling of different modes takes
place depending on the actual range of parameters including the components of
the wavenumber.  These results are not easily portable to the flux tube case
because of the geometrical differences. Nevertheless, we confirm the close
``cooperation'' of the buoyancy and shearing forces and following FT'95 apply
the unified name: ``the Parker-shearing instability''. It is however more
appropriate to use the name ``Parker instability'' in some contexts.

{The magnetic shearing (Balbus-Hawley) instability has been discussed in
connection with the buoyancy and dynamo action by Brandenburg et al.
(1995). The authors simulate the non linear evolution of three-dimensional
magnetized Keplerian shear flow. Their system acts like  dynamo that generates
its own turbulence via the mentioned instabilities. They observe signatures of
buoyancy as well as the tendency of magnetic field to form intermittent
structures.}

The plan of the paper is as follows:
In Section 2 we present a general formalism for the description of the
flux tube dynamics in galactic discs using the thin flux tube approximation.
In Section 3 we perform a linear stability analysis and discuss the relations
between the flux tube and the continuous approaches to the Parker-shearing
instability.
We calculate first the dynamo coefficients of the galactic flux tube dynamo
for the case without shear.  Then we take into account differential forces due
to the galactic axisymmetric shear and the non axisymmetric density waves as
well.  In Section 4 we demonstrate relations between our theoretical and the
recent observational results and propose a new dynamo model
based on the Parker-shearing instability in galactic discs with density waves.

\section{Magnetic flux tubes in galactic discs -- basic
equations and properties}

We are going to apply the thin, slender  flux tube approximation which according to
Spruit (1981), Spruit and Van Ballegooijen (1982) means that the tube is
assumed to be thin compared with
both the scale height $\Lambda$ of vertical density stratification and with the
scale
$L$ along the tube on which the direction of its path changes.
In the foregoing considerations we shall follow the formulation of
Moreno-Insertis (1986) with some modifications related to the
specifics of the galactic disc.
We will neglect
the details of the shape of the cross section of the flux tube and assume that
it is circular.  We will also assume that the tube is not twisted. The assumptions
of the concept of the the thin, slender flux tube are extensively discussed
 by Schramkowski and Achterberg (1993).

\subsection{Geometry of the thin flux tube}

Let us describe the shape
of flux tube in the 3-D volume and the cartesian reference frame rotating with the disc,
by means of 3 functions
\begin{eqnarray}
x&\equiv&x(t,m),\non\\
y&\equiv&y(t,m),\\
z&\equiv&z(t,m)\non
\end{eqnarray}
of time $t$ and a ``spacelike'' parameter $m$ -- the mass measured along the
tube. The lagrangian vectors
\begin{eqnarray}
\bm{v}=\left(\left(\pder{x}{t}{}\right)_m,
\left(\pder{y}{t}{}\right)_m, \left(\pder{z}{t}{}\right)_m\right),\\
\bm{L}=\left(\left(\pder{x}{m}{}\right)_t,
\left(\pder{y}{m}{}\right)_t, \left(\pder{z}{m}{}\right)_t\right),
\end{eqnarray}
are respectively a velocity vector and a tangent vector. Let us specify the
parameter $m$ by the requirement that the mass flow across points $m=\const$
vanishes, what means that we are going to use the lagrangian description.
Having a freedom to rescale the parameter $m$ we can chose the normalization
\begin{equation}
dm \equiv \xi ds, \label{m2s}
\end{equation}
where $\xi$ and $ds$ are respectively a mass per unit length and a
length of an element of the flux tube. Since
\begin{eqnarray}
ds^2 & \equiv & dx^2+ dy^2+dz^2\non\\
     & = & \left(\left(\pder{x}{m}{}\right)^2
                +\left(\pder{y}{m}{}\right)^2
                +\left(\pder{x}{m}{}\right)^2 \right) dm^2
\end{eqnarray}
then
\begin{equation}
\mid\bm{L}\mid = \fr{1}{\xi}
\end{equation}
The unit tangent vector $\bm{l}$ is given by
\begin{equation}
\bm{l} \equiv \xi \bm{L}.
\end{equation}
Now let us define the curvature vector
\begin{equation}
\bm{K} \equiv \pder{\bm{l}}{s}{}
\end{equation}
%
%In practice it will be more convenient to compute $\bm{K}$ with the aid of the
%formula
%\begin{equation}
%\bm{K} = \xi^2 \left(\pder{\bm{L}}{m}{}
%-\xi^2 \left(\pder{\bm{L}}{m}{} \cdot \bm{L}\right) \bm{L}\right)\label{curv}
%\end{equation}
%
%
Then the curvature radius is given by
\begin{equation}
\fr{1}{R}\equiv K \equiv \mid\bm{K}\mid = \xi \mid\pder{\bm{l}}{m}{}\mid
\end{equation}
and
\begin{equation}
\bm{n}\equiv \fr{\bm{K}}{K}
\end{equation}
is a unit normal vector (principal normal) parallel to the curvature vector
$\bm{K}$.  In the following considerations we will need to decompose vectors
as eg.  gravitational acceleration $\bm{g}$ into components parallel and
perpendicular to the unit tangent vector $\bm{l}$

\begin{eqnarray}
\bm{g}_{\paral} &\equiv& \left(\bm{g} \cdot \bm{l}\right)\bm{l},\\
\bm{g}_{\perp}  &\equiv& \left(\bm{l} \cross \bm{g} \right)\cross \bm{l}.
\end{eqnarray}

\subsection{MHD equations for a thin flux tube}
In order to describe the dynamics of magnetic flux tubes in a rotating
reference frame of galactic discs we shall start with MHD equations in the
following form

\begin{equation}
\pder{\rho}{t}{} + {\rm div} (\rho \bm{v}) = 0,\label{cont}
\end{equation}

\begin{eqnarray}
\lefteqn{\rho\left(\pder{\bm{v}}{t}{}+(\bm{v}\cdot\nabla)\bm{v}\right) =}
\label{eqofmot} \\
&&-\nabla p + \rho\bm{g}+\fr{1}{4\pi} \bm{j}\cross\bm{B}
 - 2\rho \, \bm{\Omega}\times\bm{v} + \bm{F}_{\rm diff}, \nonumber
\end{eqnarray}

\begin{equation}
\pder{\bm{B}}{t}{} = {\rm rot}(\bm{v}\times\bm{B})
- {\rm rot}\left(\fr{1}{4 \pi \sigma} {\rm rot} \bm{B}\right),\label{ind}
\end{equation}

\begin{equation}
{\rm div} \bm{B} = 0,\label{divfree}
\end{equation}
where we have used traditional assignments for physical variables.  The terms
on the rhs.  of the equation of motion (\ref{eqofmot}) related to rotation need
an additional explanation.  The $-2\rho\bm{\Omega}\times\bm{v}$ is the Coriolis
force.  The additional term $\bm{F}_{\rm diff}$ is the differential force due
to counteracting effects of centrifugal force and the radial component of
gravitation in the shearing sheet approximation (see eg.  FT'94 and references
therein). The differential force will be introduced in the linearized form in
the section 3. The role of the differential force due to
the galactic dynamics is discussed in more details by FT'94.
In addition to the above system of MHD equations we should specify additional
relations which are typical for the galactic disc.  Our intention is to define
a general setup for the Parker-shearing instability of magnetic flux tubes
which is
as close as possible to the traditional continuous 3D distribution of galactic
magnetic fields discussed recently in FT'94 and FT'95.  This will allow us to
show that the flux tube stability properties are physically very similar to
stability properties of the mentioned continuous distribution of interstellar
magnetic fields.

Following Parker (1966, 1967a,b) we shall assume the following isothermal equation of
state for interstellar gas
\begin{equation}
p_g = u^2 \rho,\label{gaslaw}
\end{equation}
with the sound speed of gas $u$ independent of height in the galactic disc.

The magnetic field and cosmic ray pressures in the unperturbed state are
traditionally assumed to be  proportional to the gas pressure
\begin{eqnarray}
p_{mag} \equiv \alpha p_g, && p_{cr} \equiv \beta p_g,
\end{eqnarray}
where $\alpha$ and $\beta$ are constants order of 1.

The vertical equilibrium of the composition of gas, magnetic field and cosmic
rays in the vertical gravitational field $\bm{g}=-{\rm sign(z)}|g|\bm{e}_z$ is
described by the equation
\begin{equation}
\frac{d}{dz}\left(p_g + p_{mag} + p_{cr}\right) = \rho(z) g_z
\end{equation}
If we assume for simplicity that $g$ is a constant, then
\begin{equation}
\rho_e (z) =\rho_e (0) \exp\left(-\fr{|z|}{\Lambda}\right),\label{denstr}
\end{equation}
where $\Lambda$ is the scale height of the disc given by Parker (1966)
\begin{equation}
\Lambda = (1+\alpha+\beta)\frac{u^2}{|g|}.
\end{equation}

In the following we shall define our model of galactic flux tubes, which is in
fact slightly different with respect to that proposed in HL'93.
The present model assumes that the flux tubes
\begin{enumerate}
\item are composed of magnetic field, ionized gas and cosmic
rays,
\item can execute motions which are almost independent on the ambient flux
tubes (eg. a particular flux tube can rise due to the Parker instability, while
the neighboring flux tubes are temporarily in a kind of dynamical
equilibrium).
\item are distributed in space with the filling factor close to 1,
\item are initially azimuthal. This simplifying assumption will
allow us to construct the equilibrium states necessary for the
linear stability analysis.
\item Collisions of the flux tubes are temporarily ignored. They can lead to
the magnetic reconnection processes and some reorganization of the magnetic
field structure, but this topic will be discussed elsewhere.
\end{enumerate}
Then, we can introduce a division between quantities specific to the flux tube
interior $(i)$ and its exterior $(e)$. Taking into account various component of
pressure we can postulate the following magneto--hydrostatic balance condition
\begin{equation}
p_{ig} + p_{im} + p_{icr} = p_{eg} + p_{em} + p_{ecr}, \label{hydrbal}
\end{equation}
stating that the total internal and external pressures are the same at the
cylindrical flux tube boundary and the aerodynamic drag force.  The aerodynamic
drag force will be considered for completeness in our set of equations, however
it will not be taken into account in the solutions of the present paper.  We
shall operate within a frame of linear stability analysis, so the aerodynamic
drag force, quadratic in velocities, will be neglected.  For the drag force
$\bm{F}_D$ we will adopt the formula

\begin{equation}
\bm{F}_D \equiv
-\fr{C_D \rho_e}{\pi r} \mid \bm{v}_{\perp}\mid \bm{v}_{\perp},
\end{equation}
where $\bm{F'}_D$ is the drag force per unit length (Sch\"ussler 1977, Stella
\& Rosner 1984).  $C_D$ is a drag coefficient which, over a wide range of
physical regimes characterized by subsonic motions and for Reynolds numbers
lying between 30 and $10^5$ (Goldstein 1938) is estimated to be $\sim 1 - 10$.

Now we are  going to reduce the equations to the convenient form for describing
the flux tube dynamics.
Combining the continuity equation (\ref{cont}), the induction
equation (\ref{ind}) and assuming infinite electrical conductivity one obtains
\begin{equation}
\fr{D}{Dt}\left(\fr{\bm{B}}{\rho}\right)=
\left(\fr{\bm{B}}{\rho}\cdot\nabla\right)\bm{v}
\end{equation}
Substituting $ B \equiv \Phi/\Sigma$ and $ \rho \equiv \xi/\Sigma$, where $\Phi$ is the
flux across the flux tube, $\Sigma$ is the cross section area, we can write
\begin{equation}
\fr{\bm{B}}{\rho} = \fr{\phi}{\xi} \bm{l} = \phi \bm{L},
\end{equation}
which leads finally to the equation
\begin{equation}
\cder{\bm{L}}{t}{} = (\bm{\tau} \cdot \nabla)\bm{v} = \pder{\bm{v}}{m}{}
\end{equation}
In the absence of Coriolis force the equation of motion
(\ref{eqofmot}) can be decomposed into two components:  parallel and
perpendicular to the tube axis
\begin{equation}
\left(\cder{\bm{v}}{t}{}\right)_{\paral} =
-\fr{\xi}{\rho_i}\pder{p_i}{m}{}\bm{l} +\bm{g}_{\paral}+
\fr{\bm{F}_{diff\paral}}{\rho_i}
\end{equation}

\begin{equation}
\left(\cder{\bm{v}}{t}{}\right)_{\perp} =
\fr{B^2}{4\pi\rho_i} \bm{K} + \fr{\rho_i-\rho_e}{\rho_i} \bm{g}_{\perp} +
\fr{\bm{F}_D}{\rho_i}+\fr{\bm{F}_{diff\perp}}{\rho_i},
\end{equation}
where $\bm{F}_{diff}$ is the differential force due to the galactic
dynamics, which will be introduced in more details later on.
The presence of Coriolis acceleration
\begin{equation}
\bm{a} \equiv -2\bm{\Omega}\times \bm{v}
\end{equation}
couples the parallel and perpendicular components of the equation of motion.

Let us assume that the flux tubes are aligned horizontally
in the initial state and assign
$p_{cr}(z0)=\beta u^2\rho_e(z_0)$ to the cosmic ray pressure specific for the
height $z_0$ in the galactic disc. Following Shu (1974) we assume that the
pressure gradient of the cosmic ray gas is orthogonal to the magnetic field
lines.  Then, during evolution of the particular tube the internal cosmic ray
pressure remains constant and equal to $p_{cr}(z0)$.  Thus, we can rewrite the
magneto-hydrostatic balance condition (\ref{hydrbal}) as follows

\begin{eqnarray}
\lefteqn{u^2\rho_i(t,m) + \fr{B_i^2(t,m)}{8\pi} + \beta u^2\rho_e(z_0)=}\\
&& \makebox[4cm]{} u^2\rho_e(z(t,m))(1+\alpha+\beta). \nonumber
\end{eqnarray}
Due to the magnetic flux conservation we can rewrite the
hydrostatic balance condition in the form
\begin{equation}
\rho_i =\tilde{\rho}_e - \left(\fr{\Phi}{u}\right)^2 \fr{1}{8\pi\Sigma^2},
\end{equation}
where we denoted
\begin{equation}
\tilde{\rho}_e \equiv \rho_e(z)(1+\alpha+\beta)-\beta\rho_e(z_0)
\end{equation}
and $\Phi^2 \equiv B_{i0}^2 \Sigma^2_0$ with $B_{i0}^2 \equiv 8\pi\alpha u^2
\rho_e(z_0)$.
Since $\xi=\rho_i \Sigma$ then the equation relating the two
unknowns $\xi$ and $\Sigma$ is
\begin{equation}
\tilde{\rho}_e \Sigma^2 - \Sigma \xi - \alpha\rho_e(z_0)\Sigma^2_0
= 0. \label{xisigma}
\end{equation}
We find that
\begin{equation}
\Sigma = \fr{\xi + \sqrt{\xi^2 + 4\alpha\rho_e(z_0)\tilde{\rho}_e\Sigma^2_0}}
{2 \tilde{\rho}_e}. \label{sigma}
\end{equation}
The final form of the system of equations is as follows
\begin{eqnarray}
\lefteqn{\cder{\bm{x}}{t}{}=\bm{v},}\label{Dxt}\\
\lefteqn{\cder{\bm{L}}{t}{}=\pder{\bm{v}}{m}{},}\label{DLt}\\
\lefteqn{\cder{\bm{v}}{t}{}=
-u^2 \Sigma\pder{\rho_i}{m}{}\bm{l} +\bm{g}_{\paral}
+\fr{B^2}{4\pi\rho_i} \bm{K} + \fr{\rho_i-\rho_e}{\rho_i}
\bm{g}_{\perp}} \label{tubemot} \\
&&- 2\bm{\Omega}\times \bm{v} + \fr{\bm{F}_{D}}{\rho_i}
+ \fr{\bm{F}_{diff}}{\rho_i}.\nonumber
\end{eqnarray}

\subsection{Basic parameters and physical units}

Let us now specify the values of basic physical parameters involved in our
considerations. Following Parker (1979) p. 806 we shall use parameters
typical for the Milky Way as an example and take
\begin{equation}
g\simeq 2\cdot 10^{-9} {\rm cm \  s^{-2}},
\end{equation}
\begin{equation}
\rho\simeq1.6\cdot 10^{-24} {\rm g \ cm^{-3}},
\end{equation}
which is equivalent to 1 H-atom cm$^{-3}$.
The conventional value of the magnetic field strength is
\begin{equation}
B \simeq 3\cdot 10^{-6}{\rm G}
\end{equation}
which leads to a magnetic pressure $p_m=B^2/8\pi=0.4\cdot 10^{-12}{\rm dyn \
cm}^{-2}$.
For comparison the typical value of the cosmic ray pressure is
\begin{equation}
p_{cr} \simeq 0.5\cdot 10^{-12} {\rm dyn \ cm^{-2}}
\end{equation}
and a gas pressure
\begin{equation}
p_{g} = \rho u^2 \simeq 0.4\cdot 10^{-12} {\rm dyn \ cm^{-2}},
\end{equation}
where $u \simeq7 {\rm km \ s}^{-1}$ is the thermal velocity of gas.
It is worthwhile to notice that Parker considers
the gas pressure due to the both thermal and turbulent gas motions.
Since the above estimations of gas, magnetic field and cosmic ray pressures
give almost identical contributions of all the 3 components then
\begin{equation}
\alpha \simeq \beta \simeq 1
\end{equation}
and the resulting vertical scale height is
\begin{equation}
\Lambda \simeq 2.3 \cdot 10^2 {\rm pc}.
\end{equation}

In the following considerations we shall use convenient units: 1 pc as a unit
of distance and 1 Myr as a unit of time.
The unit of velocity is
\begin{equation}
1\pc\Myr^{-1} \simeq 0.98 \km\s^{-1}
\end{equation}
and the unit of acceleration is
\begin{equation}
1\pc\Myr^{-2} \simeq 3.09\cdot10^{-14} \km\s^{-2}.
\end{equation}
In these units
\begin{equation}
g \simeq 0.65 \pc\Myr^{-2}
\end{equation}
and
\begin{equation}
\Omega \simeq 3.2 \cdot 10^{-2} \Myr^{-1}.
\end{equation}

\section{The linear stability analysis}

\subsection{Equations}

With the aim of performing the linear stability analysis we decompose all the
quantities in the 0-th and 1-st order parts. Let us start with
\begin{equation}
m = m_0 + m_1(m_0,t)
\end{equation}
which leads to
\begin{equation}
\pder{}{m}{} = \left(1-\pder{m_1(m_0,t)}{m_0}{}\right)\pder{}{m_0}{}.
\end{equation}
For the flux tube which is extended along the $x$-coordinate in the unperturbed
state, the coordinates of a given flux tube element in the linear approximation
are
\begin{eqnarray}
x&=&\fr{m_0}{\xi_0} + x_1\\
y&=&y_1,\\
z&=&z_1.
\end{eqnarray}
The components of the tangent vector in the linear approximation are
\begin{eqnarray}
L_{x0}&=&\fr{1}{\xi_0}\left(1-\pder{m_1}{m_0}{}\right)
+\pder{x_1}{m_0}{},\\
L_{y_0}&=&\pder{y_1}{m_0}{},\\
L_{z_0}&=&\pder{z_1}{m_0}{}.
\end{eqnarray}
The modulus of the tangent vector, the unit tangent vector and the curvature
vector are respectively
\begin{eqnarray}
\mid\bm{L}\mid &=&\fr{1}{\xi_0}\left(1-\pder{m_1}{m_0}{}
                  +\xi_0\pder{x_1}{m_0}{}\right),\\
\bm{l}&=&\left[1,\xi_0\pder{y_1}{m_0}{},\xi_0\pder{z_1}{m_0}{}\right],\\
\bm{K}&=&\xi^2_0\left[0,\pder{y_1}{m_0}{2}, \pder{z_1}{m_0}{2}\right].
\end{eqnarray}
Using the equation (\ref{DLt}), one can find that:
\begin{equation}
\pder{}{t}{}\pder{m_1}{m_0}{} = 0,
\end{equation}
which implies that
\begin{equation}
m_1 \equiv 0.
\end{equation}
The mass per unit length of the tube in the linear approximation is
\begin{equation}
\xi = \xi_0 \left( 1 - \xi_0 \pder{x_1}{m_0}{} \right).
\end{equation}
Now we calculate $\rho_{i1}$ in the following sequence
\begin{equation}
\xi_1= - \xi^2_0 \pder{x_1}{m_0}{},
\end{equation}

\begin{equation}
\tilde{\rho}_{e1} = -\rho_e(z_0) \fr{|g|}{u^2} z_1,
\end{equation}

\begin{equation}
\Sigma_1 = \fr{\Sigma_0}{2\alpha+1}   \left(\fr{\xi_1}{\xi_0}
+\fr{|g|}{u^2} z_1\right),
\end{equation}
\begin{equation}
\rho_{i1}= \fr{2\alpha}{2\alpha+1} \fr{\xi_1}{\Sigma_0}
 - \fr{1}{2\alpha+1}\fr{\xi_0}{\Sigma_0}
   \fr{|g|}{u^2} z_1
\end{equation}
Let us derive some coefficients appearing in the equation of motion
(\ref{tubemot})
\begin{equation}
\left(\fr{\rho_i-\rho_e}{\rho_i}\right)_1 =
\left(\fr{-1}{2\alpha+1}\fr{|g|}{u^2}-\fr{1}{\Lambda}\right) z_1
+\fr{2\alpha}{2\alpha+1}\fr{\xi_1}{\xi_0},
\end{equation}
\begin{equation}
\left(\fr{B^2}{4\pi\rho_i}\bm{K} \right)_1 = v_A^2 \bm{K}_1,
\end{equation}
where $v_A^2 = 2\alpha u^2$.
The equation of tube motion (\ref{tubemot}) after linearization is
\begin{eqnarray}
\pder{\bm{v}_1}{t}{}\b&=&
\b \fr{v_A^2}{1+2\alpha}\left(\pder{x_1}{s}{2}
   - \fr{(1+\alpha+\beta)}{\Lambda}\pder{z_1}{s}{}
\right) \bm{l}_0 \\
 &+&\fr{1+\alpha+\beta}{1+2\alpha}\left(
\fr{v_A^2}{\Lambda}\pder{x_1}{s}{} - (\alpha-\beta) \fr{u^2}{\Lambda} z_1
\right)\bm{g}\nonumber\\
&+& v_A^2 \bm{K}_1  -2\bm{\Omega}\cross\bm{v}_1 + \bm{a}_{diff},\nonumber
\label{eqmots}
\end{eqnarray}
where $s$ is the length parameter of the tube defined by (\ref{m2s})
and
\begin{eqnarray}
\bm{g} &=& [0,0,g],\\
\bm{l}_0&=&[1,0,0],\\
\bm{K}_1&=&\left[0,\pder{y_1}{s}{2},\pder{z_1}{s}{2}\right],\\
-2\bm{\Omega}\cross\bm{v}_1&=&\left[2\Omega\pder{y_1}{t}{},
-2\Omega\pder{x_1}{t}{},0\right]
\end{eqnarray}

The linear effect of an axisymmetric differential rotation can be incorporated
in a way similar to  FT'94.
The differential force is radial in direction and is proportional to the radial
displacement  in the shearing sheet approximation.   There is an obvious
difference between the 3D continuous case and the flux tube approach.
In the continuous case the differential rotation influences also the radial
wavenumber which varies linearly as a function of time (FT'94).
In the flux tube approach the wavevector of the perturbation is always parallel
to the tangent vector and
this effect is absent since we consider (until now) only azimuthal flux tubes
which are initially localized radially and vertically.

The acceleration operating on an element of the flux tube in the shearing
sheet approximation is

 \begin{equation}
\bm{a}_{diff} \equiv -4 A \Omega y_1 {\bm e}_y
\end{equation}
where
\begin{equation}
A= \frac{R_0}{2} \frac{d\Omega}{dR}
\end{equation}
is the Oort constant ($A/\Omega = -1/2$ for flat
rotation and $-3/4$ for the Keplerian rotation).

\subsection{The linear solution}

In a case of vanishing rotation and shear we assume the harmonic
perturbation on a magnetic flux tube placed in the coordinate system $(x,z)$
in the form
\begin{equation}
\left[\matrix{x_1 \cr z_1}\right] =
\left[\matrix{X_1 \cr Z_1}\right] \exp i(k s -\omega t) + {\rm c.c.}
\end{equation}
where $s$ is a length parameter measuring position along the flux tube, $k$ is
the wavenumber, $x$ and $z$ are the azimuthal and vertical coordinates
respectively.  The linearized equations describing evolution of harmonic
perturbations on the flux tube lead to the linear dispersion relation

\begin{equation}
\det\b\left[\matrix{\frac{v_A^2 k^2}{1+2\alpha} -\omega^2&
\b\frac{1+\alpha+\beta}{1+2\alpha}\frac{v_A^2}{\Lambda} i k \cr
\frac{1+\alpha+\beta}{1+2\alpha}\frac{v_A^2}{\Lambda} (-i k) &
\b\frac{(\alpha-\beta)(1+\alpha+\beta)}{1+2\alpha}
\frac{u^2}{\Lambda^2}+v_A^2k^2 -\omega^2
}\b\right]\! =\! 0,
\end{equation}
where $u$ and $v_A$ are the thermal and Alfven speed respectively, and
$\Lambda = (1+\alpha+\beta) u^2/\mid g \mid$ is the vertical scaleheight.
After the substitution of the dimensionless:  azimuthal wavenumber $\bar{k} = k
\Lambda$ and complex frequency $\tilde{\omega} = \omega \Lambda / v_A$, the
above dispersion relation reduces exactly to the form (A1) in FT'94,
which represents a case of
continuous 3D distribution of magnetic field
with vanishing vertical wavenumber $k_z = 0$ and the radial wavenumber $k_r =
\infty$.  This means that our flux tube approach is analogous to
the above limiting case of continuous distribution of magnetic field.

Nevertheless, as soon as we take into account rotation (see below), the analogy
becomes less clear.  Our dispersion relation becomes of third order in
$\omega^2$, and the marginal stability point $k_{marg}$ (which can be determined
after the
substitution of $\omega = 0$ to the dispersion relation)  is the same as for no
rotation.  In the continuous case of FT'94, the marginal instability point is
the same as for no rotation ($k_P$) for the radial wavenumber $k_r\to\infty$,
and is smaller ($k_Q<k_P$) for $k_r=0$.  But in the limit $k_r\to\infty$, the
dispersion relation of FT'94 is independent of the rotation
($\Delta(\omega^2)=0$), and of second order in $\omega^2$ (their eq.  A1).  The
main difference with our approach is that the optimal wavenumber $k$ leading to
the maximum growth rate is independent of the rotation in the continuous case,
whereas it depends on the rotation in the flux tube approach according to our
Fig.~1.  This is mathematically different, but physically acceptable, since
the trends observed in our Fig.~1 and in Fig.~1.  of FT'94 are very similar.
Thus, we can conclude that the slender flux tube model, in addition to its
observational justification, has the advantage to reduce the fully
three-dimensional MHD equations to a curvilinear one-dimensional formalism.
Similar conclusion has been already made by Schramkowski and Torkelson
(1996) in the context of accretion discs.

In the presence of nonvanishing galactic rotation one should take into account
also the radial coordinate $y$. In this case the linear harmonic perturbations
are assumed to be
\begin{equation}
\left[\matrix{x_1 \cr y_1 \cr z_1}\right] \equiv
\left[\matrix{X_1 \cr Y_1\cr Z_1}\right] \exp i(k s -\omega t) + {\rm c.c.}
\end{equation}
We can write the  dispersion relation in the form

\begin{eqnarray}
\lefteqn{\det\b\left[\matrix{\frac{v_A^2 k^2}{1+2\alpha} -\omega^2 &
\b\b\b\b\b\b 2 i \Omega\omega &
\b\b\b\frac{1+\alpha+\beta}{1+2\alpha}\frac{v_A^2}{\Lambda} i k \cr
\b -2 i \Omega\omega & \b\b\b v_A^2 k^2\b +4A\Omega- \omega^2 &\b\b\b\b\b\b 0
\cr \b\frac{1+\alpha+\beta}{1+2\alpha}\frac{v_A^2}{\Lambda} (-i k) &\b\b\b 0 &
\b\b\b\b\b\b\frac{(\alpha-\beta)(1+\alpha+\beta)}{1+2\alpha}
\frac{u^2}{\Lambda^2}\!+\! v_A^2k^2\b -\omega^2
}\b\right]}\nonumber\\
&& = 0 \label{disprel}
\end{eqnarray}
If $\omega$ is a solution of the above equation representing the Parker mode
for given $k$, then eigenvectors are given by the following relations between
constants $X_1$, $Y_1$ and $Z_1$

\begin{eqnarray}
\frac{X_1}{Z_1} &=& \frac{
\frac{(\alpha-\beta)(1+\alpha+\beta)}{1+2\alpha}\frac{u^2}{\Lambda^2}
+ v_A^2 k^2 - \omega^2}
{\frac{1+\alpha+\beta}{1+2\alpha}\frac{v_A^2}{\Lambda}(i k)}\label{X2Z}\\
\frac{Y_1}{Z_1} &=& \frac{2 i \Omega \omega}{v_A^2 k^2 +4 A \Omega-\omega^2}
\frac{X_1}{Z_1}\label{Y2Z}
\end{eqnarray}

In the Lagrangian description, the dynamo coefficients $\alpha_d$ and
$\eta_d$ can be calculated using the method applied by Ferriz-Mass et. al
(1994). The expression for $\alpha_d$ is
\begin{equation}
\alpha_d \equiv \frac{\langle \bm{v} \times \bm{b}\rangle_x}{\langle \bm{B}
\rangle_x} =\langle \dot{\bm{r}} \times
\acute{\bm{r}}\rangle_x,\label{alphaddef} \end{equation}
where $\bm{b}$ is the first order perturbation of magnetic
field, $\bm{v}=\dot{\bm{r}}$ is the velocity of a lagrangian element of the
flux tube and $\acute{\bm{r}} = \partial \bm{r}/\partial s$ is the tangent
vector. The angle braces $\langle \cdots \rangle$ stand for space averaging,
which in our case is related to integration over the parameter $s$.
Similarly, the expression for diffusivity is
\begin{equation}
\eta_d \equiv \langle \dot{\bm{r}} \cdot \bm{r} \rangle \label{etaddef}.
\end{equation}
Assuming only vertical diffusion for simplicity we get
\begin{equation}
\eta_d = \langle \dot{r}_z r_z \rangle.
\end{equation}
{This simplification is valid if horizontal displacements of the flux tube
are smaller than the vertical ones. We shall indicate a case which does not
fulfill this requirement.}

After substitution of the linear solution one obtains
\begin{eqnarray}
\alpha_d &=&  \omega_i k Z_1(t)^2
\Im \left(\frac{Y_1}{Z_1}\right)\label{alphad}\\
\eta_d   &=& \frac{\omega_i}{2} Z_1(t)^2 \label{etad}
\end{eqnarray}
where $Z_1(t) = Z_1  \exp(\omega_i t)$ is the amplitude of vertical
displacement of the flux tube.
{The neglected horizontal diffusivity $\eta_{dh} = \fr{\omega_i}{2}
Y_1(t)^2$ can
be easily estimated basing on vertical diffusivity since $\eta_{dh} = \eta_{dv}
|Y_1/Z_1|^2$.}

The magnetic Reynolds numbers and the dynamo number are defined as
\begin{eqnarray}
R_\alpha &\equiv& \frac{\alpha_d \Lambda}{\eta_d}\label{Ralpha}\\
R_\omega &\equiv& \frac{\Omega \Lambda^2}{\eta_d}\label{Romega}\\
D        &\equiv& R_\alpha R_\omega\label{D}
\end{eqnarray}

{The coefficients given by the relations (\ref{alphad})--(\ref{D}) depend
on all the involved parameters, especially on the vertical density gradient
and the magnitude of shear implicitly,
via the solutions of the dispersion relation (\ref{disprel}), $\omega
= \omega(k)$ and the relations (\ref{X2Z}) and (\ref{Y2Z}) between
components of the eigenvectors. The most relevant quantities contributing to
$\alpha_d$ and $\eta_d$ will be presented in a series of forthcoming figures.}

\subsection{The numerical results}

Fig.~1 shows the growth rate vs.  wavenumber for $\alpha=\beta=1$, and other
parameters are described in the subsection 2.3.  The full line is for the case without
rotation, the dotted line is for the rotation $\Omega= \Omega_G= 10^{-15} {\rm
s}^{-1} = 0.03 {\rm Myr}^{-1}$, the dashed line is for $\Omega=2\Omega_G$, and
dashed-dotted line is for $\Omega=3\Omega_G$.  As is well known, increasing the
rotation frequency $\Omega$ dimishes the growth rate.

%%%%%%%%%%%%%%%%%%%%%%%%%%%%  FIG.1 %%%%%%%%%%%%%%%%%%%%%%%%%%%%%%%%
\begin{figure}
\epsfxsize=\hsize \epsfbox{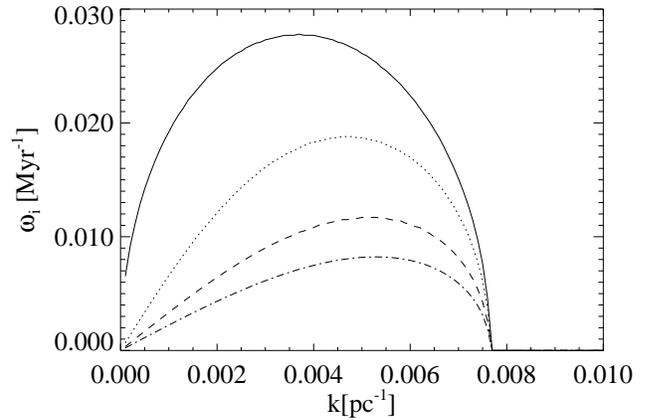}
\caption[]{
The dependence of growth rate $\omega_i$ on wavenumber $k$
for different values of the galactic rotation frequency $\Omega$
}
\end{figure}
%%%%%%%%%%%%%%%%%%%%%%%% END OF FIG. 1 %%%%%%%%%%%%%%%%%%%%%%%%%%%%%

It is worth noting, however that the ratio of amplitudes $Y_1/Z_1$ of radial to
 vertical displacements which measures the magnitude of cyclonic
deformation of the flux tube increases with rotation $\Omega$ (see Fig.~2, where
dotted, dashed and dotted-dashed curves represent the same values of $\Omega$
as in Fig.~1).  This ratio contributes to the $\alpha_d$ coefficient in (\ref{alphad}).
In addition, in the limit of $k \rightarrow 0$,
$Y_1/Z_1 \rightarrow 1$ for all values of the rotation frequency.

%%%%%%%%%%%%%%%%%%%%%%%%%%%  FIG.2 %%%%%%%%%%%%%%%%%%%%%%%%%%%%%%%%
\begin{figure}
\epsfxsize=\hsize \epsfbox{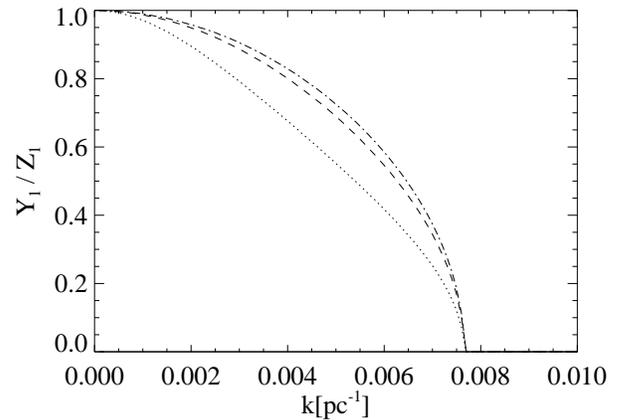}
\caption[]{
The dependence of the ratio of amplitudes $Y_1/Z_1$  on $k$
for the same values of values of $\Omega$ as in Fig.~1.
}
\end{figure}
%%%%%%%%%%%%%%%%%%%%%%%% END OF FIG. 2 %%%%%%%%%%%%%%%%%%%%%%%%%%%%%

The dependence of $\alpha_d$ on $k$ is shown in Fig.~3 for $\Omega=\Omega_G/2$
(continuous line), $\Omega=\Omega_G$ (dotted line), $\Omega=2\Omega_G$ (dashed
line) and $\Omega=3 \Omega_G$ (dashed-dotted line).  The $\alpha_d$ coefficient
is computed with the assumption that the uppermost part of the tube has passed
the vertical distance equal to the vertical scale height $\Lambda$:  $Z(t) =
\Lambda$ for each value of $k$ (for $\alpha=\beta=1$, $u = 7 \km\s^{-1}$ and
$\mid g \mid = 2 \cdot 10^{-9} \cm\s^{-1} = 0.65  \pc\Myr^{-2}$ we have
$\Lambda = 226 \pc$). This calculation of $\alpha_d$ following Ferriz-Mas
et al. (1994) is intended only to give a first insight about the dependence
of $\alpha_d$ on $k$ and $\Omega$.  It hides, however our ignorance about:  (1)
the initial amplitude of perturbations and (2) the final state reached by a
single flux tube.  Thus, within the linear theory we are not able to average
over a statistical ensemble of flux tubes.  Fig.~3 shows that the observed value
of the galactic rotation in the solar neighborhood approximately maximizes the
magnitude of the $\alpha_d$ coefficient.
This property could be quite important since many
qualitative approaches adopt  $\alpha_d$  proportional to $\Omega$
(see eg. Ruzmaikin et al. 1988).
What we observe in our model is in contradiction with this rule.

%%%%%%%%%%%%%%%%%%%%%%%%%%%%  FIG.3 %%%%%%%%%%%%%%%%%%%%%%%%%%%%%%%%
\begin{figure}
\epsfxsize=\hsize \epsfbox{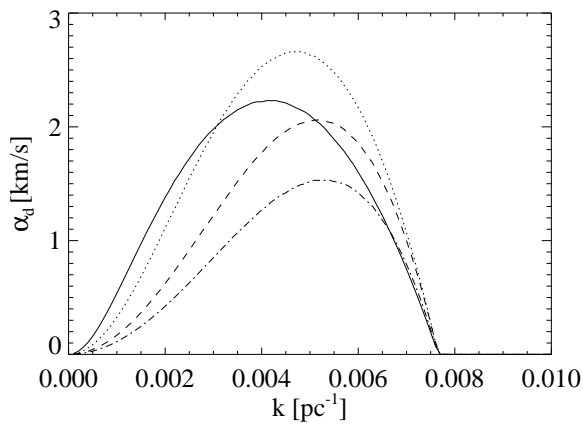}
\caption[]{
The dependence of the dynamo-$\alpha_d$ coefficient on $k$ for
a fixed vertical displacement $Z_1=\Lambda$ and
$\Omega= 1/2,1,2,3\cdot \Omega_G$.
}
\end{figure}
%%%%%%%%%%%%%%%%%%%%%%%% END OF FIG. 3 %%%%%%%%%%%%%%%%%%%%%%%%%%%%%

{It is worthwhile to notice that the sign of
$\alpha_d$ is positive for positive $z$ coordinate,
as it follows from the nature of Coriolis force which tends to slow down the
rotation of rising and expanding volumes.  The flow of gas directed outward
of the
top of the rising flux tube can be recognized as expanding in the direction
along the magnetic field lines.

This feature of our model is in contrast with the results of Brandenburg
et al.  (1995).  Strictly speaking they discuss two quantities:  the mean
helicity $<v \cdot {\rm rot} v>$ and the $\alpha_{dyn}$ defined as a
coefficient relating the azimuthal component of electromotive force with the
azimuthal mean magnetic field.  While the first quantity has a proper sign
('-' in their convention of assignments, which is consistent with the positive
sign of our $\alpha_d$), the negative sign $\alpha_{dyn}$ coefficient seems to
be in conflict with the positive sign of helicity.

We explain that the $\alpha_d$ coefficient in our model is an equivalent of the
helicity $<v \cdot {\rm rot} v>$ used by Brandenburg et al.  (1995).  On the
other
hand we are dealing with a simple model in which the dynamo properties are
derived from the dynamics of a single flux tube.  Studies of the dynamics of an
ensemble of flux tubes would help to find a reason of the discrepancy.
Moreover, we remain in the linear regime in the present paper, thus a more
detailed comparison is rather difficult.}

Let us now examine the temporal evolution of the dynamo coefficients
$\alpha_d$, $\eta_d$, $R_{\alpha}$, $R_{\omega}$ and $D$ computed within the
linear model. The results for $\alpha=\beta=1$, $\Omega= \Omega_G$ and the
maximally unstable Parker mode with $\lambda = 1330 {\rm pc}$ are
shown in Fig.~4.

It is apparent from Fig.~4 that starting from a reasonable value of the
amplitude of initial perturbation ($Z_{init}=10{\rm pc}$) the $\alpha_d$
coefficient reaches the value of $10 \km \s^{-1}$ within the galactic rotation
period
$T_G = \pi/\Omega_G \simeq 200 {\rm Myr}$.  In the same time $\eta_d$ reaches
the value of $1000 \pc^2\Myr^{-1} = 3\cdot 10^{26} \cm^2\s^{-1}$.  The two
curves have the same slope because of the same time-dependent factor $\exp (2
\omega_i t)$.  For the same reason the coefficient $R_\alpha$ is a constant
approximately equal to 1.  The coefficients $R_{\omega}$ and $D$ diminish with
time due to growing $\eta_d$ starting from the values of thousands and finishing
on 1 after the time $T_G$.  It is worth noting that we deal with an interesting
situation of growing of $\alpha_d$ and simultaneously decreasing $D$.

{In the case of an ensemble of flux tubes, which represents a more
realistic situation, the dynamo transport coefficients will result from
averaging of single flux tube contributions.  Any estimations of their mean
values has to be dependent on additional assumptions about a global flux tube
network model.  We can notice a variety of a possibilities.  The feature, which
seems to be the most decisive for the transport coefficients is the level of
coherency of the flux tube motions.  We can say that a high degree of coherency
(at least in some finite size domains) should favour higher values of the
mean transport coefficients, making them more comparable to the single flux
tube
values.  In the case of low degree of coherency the system contains flux tubes
more and less advanced in their buoyant rise, so their ensemble averages will
represent some rather moderate values.  In addition, the lower degree of
coherency of flux tube motions should result in more frequent collisions
between flux tubes, which in turn should limit the transport effects.  Our
intuition is however limited to the cases of free flux tube motions, so it is
quite difficult to derive any save conclusions at the moment.  }

Fig.~3 gives the impression that the maximum of $\alpha_d$ is related to the
maximum of instability of the Parker mode. It is not true in general for an
arbitrary position in the $(\alpha,\beta)$-plane, because $\omega_i$ is not
the only factor in eq. (\ref{alphad}).  In order to demonstrate this effect, we
fix the cosmic ray pressure putting  $\beta=1$, and vary
$\alpha$. The results are shown in Fig.~5.

%%%%%%%%%%%%%%%%%%%%%%%%%%%  FIG.4 %%%%%%%%%%%%%%%%%%%%%%%%%%%%%%%%
\begin{figure}
\epsfxsize=\hsize \epsfbox{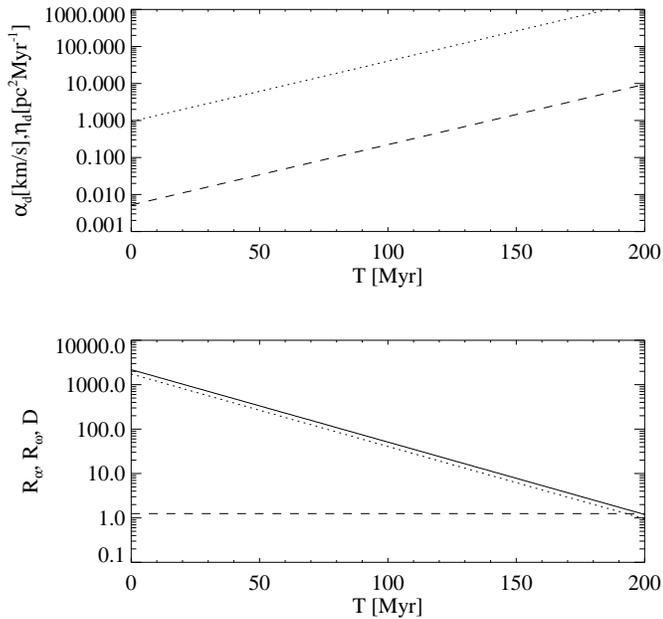}
\caption[]{
The dependence of coefficients $\alpha_d$ (dashed line) and $\eta_d$ (dotted
line) on time in the linear approximation starting from an arbitrarily assumed
initial vertical displacement $Z_{init} = 10 {\rm pc}$ is shown in the upper
panel.  The associated dynamo numbers $R_{\alpha}$ (dashed line), $R_{\omega}$
(dotted line) and $D$ (continuous line) are shown in the lower panel.

}
\end{figure}
%%%%%%%%%%%%%%%%%%%%%%%% END OF FIG. 4 %%%%%%%%%%%%%%%%%%%%%%%%%%%%%
%

%%%%%%%%%%%%%%%%%%%%%%%%%%%  FIG.5 %%%%%%%%%%%%%%%%%%%%%%%%%%%%%%%%
\begin{figure}
\epsfxsize=\hsize \epsfbox{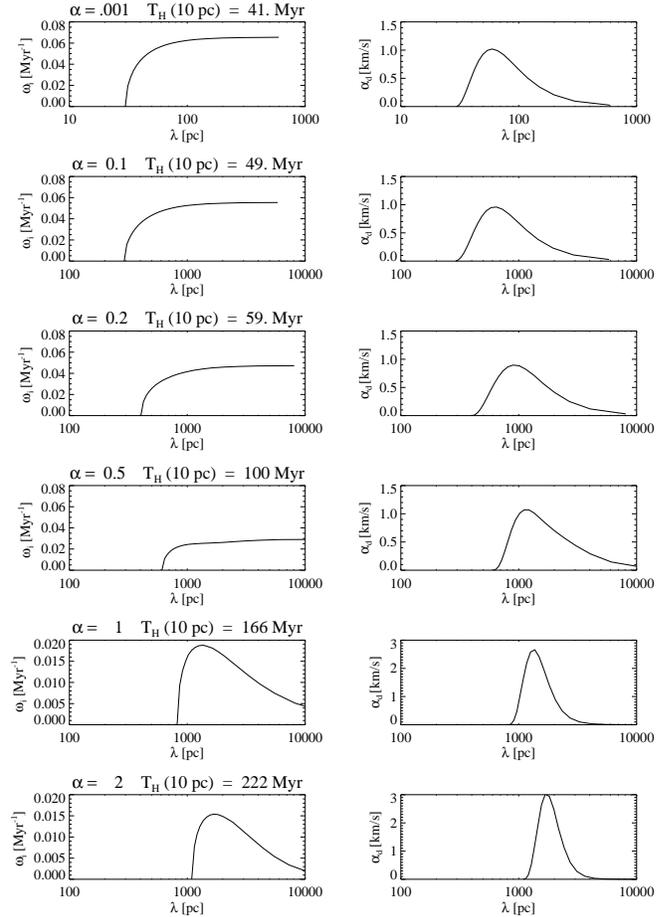}
\caption[]{
The growth rate $\omega_i$ (first column) and the coefficient $\alpha_d$
(second column) as a function of $k$ for $\beta=1$ and $\alpha$
varying.
}
\end{figure}
%%%%%%%%%%%%%%%%%%%%%%%%% END OF FIG. 5 %%%%%%%%%%%%%%%%%%%%%%%%%%%%%
%

Let us first look at the left column of graphs
which shows the growth rate for various values of $\alpha$ and $\beta=1$.
We notice that :
\begin{enumerate}
\item For $\alpha$ comparable to 1 and larger the marginal stability point
($\omega_i=0$) is placed near the wavelength $\lambda= 1000 \pc$ and the
maximum instability is attained for a bit longer wavelength (eg. $\lambda
(\omega_{max}) = 1330 \pc$ for $\alpha =1$).
\item Decreasing $\alpha$ below approximately 0.6 results in a change of
qualitative character of the curve $\omega_i(\lambda)$: the curve has no longer
a
maximum at a finite wavelength and grows monotonically with growing wavelength.
For wavelengths long with respect to the marginal wavelength the growth rate
tends asymptotically to a constant value.
\item The marginal stability point shifts to shorter and shorter wavelengths
and the maximum value of the growth rate grows while decreasing
$\alpha$. This can be simply explained since the magnetic tension
counteracting the buoyancy force is smaller for smaller $\alpha$. For a small
value of $\alpha=0.001$ (the first row) the marginal stability point is placed
at a few tens of parsecs.
\end{enumerate}
On the base of these remarks we find the maximum growth rate over a full range of
wavelengths and fixed $\alpha$, $\beta$ and assign it $\omega_{i\,max}$.
Next, with this growth rate we calculate the growth time $T_{\Lambda}
(Z_{init})$ which is defined as the time which is necessary to magnify the Parker
mode from given amplitude of vertical displacement $Z_{init}= 10 {\rm pc}$ to a
value $Z_1 = \Lambda$, the scale height of the Parker instability.

\begin{equation}
T_{\Lambda} (Z_{init}) =
\frac{1}{\omega_{i\,max}} \log \left(\frac{\Lambda}{Z_{init}}\right)
\end{equation}
The value of $T_{\Lambda} (Z_{init})$ is indicated above each row.  Now let us
point our attention on the second column of graphs, where $\alpha_d$ is
computed for  $T_{\Lambda} (Z_{init})$ as a function of $\lambda$.
Note: The same value of $T_{\Lambda} (Z_{init})$ is used for each
$\lambda$.

 The result can be commented as follows: the strongest dynamo $\alpha$-effect
(maximum of $\alpha_d$) comes from modes placed near the marginal stability
point, namely the maximum of $\alpha_d$ is due to the modes of wavelength below
100 pc for weak magnetic fields $\alpha =0.001$. Thus, if our flux tube is
perturbed with a superposition of Parker modes which have different wavelengths
and comparable initial amplitudes then we can expect a main contribution to the
total $\alpha$-effect due to the modes of wavelength $\lambda
(\max{\alpha_d})$ even if these modes do not have the maximum growth rate.

One should notice that the times $T_{\Lambda}(Z_{init})$ are very different
depending on $\alpha$ and for this reason the numerical values of
$\alpha_d$ are incomparable between rows of Fig.~5.
For the sake of comparing the rate of generation of the $\alpha_d$-effect as a
function of $\alpha$ and $\beta$, it is better to fix time. Let us take $T=50 {\rm
Myr}$.  Fig.~6. shows the values of  ${\rm Max}(\alpha_d(\lambda))$ over the
$(\alpha, \beta)$-plane. The next Fig.~7 shows the  wavelengths of the
Parker modes which give rise to these maxima.

%%%%%%%%%%%%%%%%%%%%%%%%%%%  FIG.6 %%%%%%%%%%%%%%%%%%%%%%%%%%%%%%%%
\begin{figure}
\epsfxsize=\hsize \epsfbox{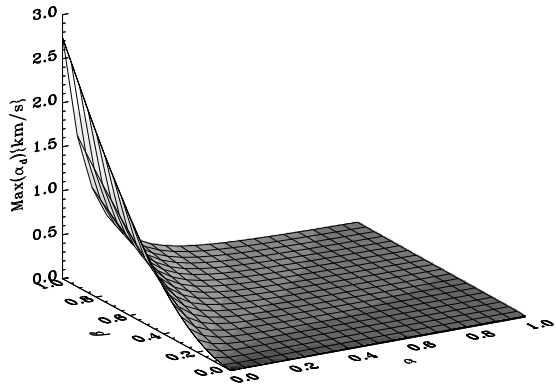}
\caption[]{
The maxima of $\alpha_d$ as a function
of $\alpha$ and $\beta=1$ and $\alpha$ varying.
}
\end{figure}
%%%%%%%%%%%%%%%%%%%%%%%%% END OF FIG. 6 %%%%%%%%%%%%%%%%%%%%%%%%%%%%%
%%
%%%%%%%%%%%%%%%%%%%%%%%%%%%  FIG.7 %%%%%%%%%%%%%%%%%%%%%%%%%%%%%%%%
\begin{figure}
\epsfxsize=\hsize \epsfbox{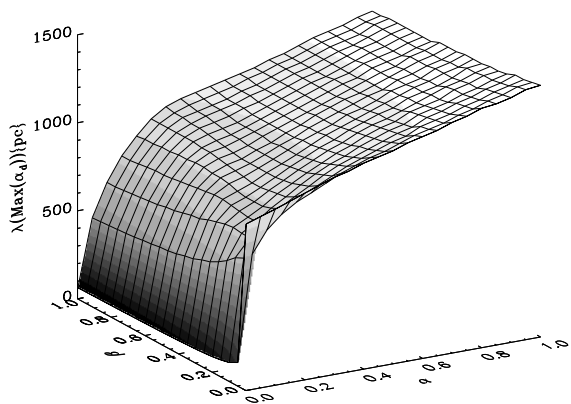}
\caption[]{
The wavelengths which give rise to the maxima of $\alpha_d$.
}
\end{figure}
%%%%%%%%%%%%%%%%%%%%%%%%% END OF FIG. 7 %%%%%%%%%%%%%%%%%%%%%%%%%%%%%
%%
At this point we should comment on an important question which arises in
relation to the above results.  The question is:  If we have an excess of
cosmic ray pressure, wouldn't the magnetic field escape from the galaxy, in an
axisymmetric way, without having time to produce a dynamo ?  For $\alpha=0.001$
the dynamo coefficient $\alpha_d$ is maximum at $\lambda<100pc$, but the
highest growth rate is for the axisymmetric interchange mode, faster than the
undulate Parker mode.

To answer this question we should note that the dominance of the axisymmetric
interchange mode results from an oversimplification of the linear stability
analysis.  It is hard to imagine the azimuthal flux tube uniformly rising in
the external medium of nonuniform density, especially in our model, in which
the flux tubes are anchored at molecular clouds (HL'93, see also Beck et
al. 1991).
The flux tubes in between molecular clouds undergo buoyant motions, but the
clouds are heavy enough to impede the long wavelength modes.
%
%regard their motion as almost independent of flux
%tubes (see, however Clifford and Elmegreen (1983), where the concept of flux
%tubes has been applied in the calculations of the cross section areas of
%molecular clouds due to the magnetohydrodynamical interactions).
%
In our linear stability analysis we treat the chaotic motions of clouds as a
source of initial perturbations.  This topic will be discussed in more details
in the forthcoming paper where we perform numerical simulations of the
nonlinear Parker-shearing instability of flux tubes.

Summarizing, we point out that even in the limit of very weak magnetic field
strengths and high cosmic ray pressure, the dominating wavelength has to be
comparable to the mean distance between clouds.  This limit of is also
favorable for the formation of molecular clouds via the Parker instability
since the typical growth rate of the Parker instability is much higher due to
the very weak magnetic tension.

{This is now evident from Fig. 6. that the cosmic rays are essentially
responsible for the strong $\alpha$-effect in the limit of weak magnetic field
(i.e. for the small $\alpha$ on the horizontal axis).
They supply buoyancy, but in contrast to the magnetic field contribution their
contribution is free of magnetic tension acting against buoyancy. This implies
the enhanced growth rate as well as the fact that the marginal
stability point is shifted toward the shorter wavelength. (The criterion for
the Parker instability determining the marginal stability point follows from
the balance between buoyancy and the magnetic tension. The magnetic tension
grows with curvature of magnetic field lines proportionally to $k^2$.)
In the result the contribution of cosmic rays is 2-folding (see the relation
(\ref{alphad})): via the growth rate and the wavenumber as well.}

\subsection{The effect of an axisymmetric differential rotation}

The differential force acts against the radial component of the magnetic
tension which depends on the value of the wavenumber $k$.  Let us fix our
attention on the the case $\alpha=\beta=1$, $\Omega = \Omega_G$ and all other
parameters like those used in Figs.  1, 2 and 3.  The solutions of the
dispersion relation (\ref{disprel}) for nonvanishing differential force
together with the ratio of amplitudes $Y_1/Z_1$ and the $\alpha_d$ coefficient
are presented in Fig.~8.

%%%%%%%%%%%%%%%%%%%%%%%%%%%%  FIG.8 %%%%%%%%%%%%%%%%%%%%%%%%%%%%%%%%
\begin{figure*}
\epsfxsize=\hsize \epsfbox{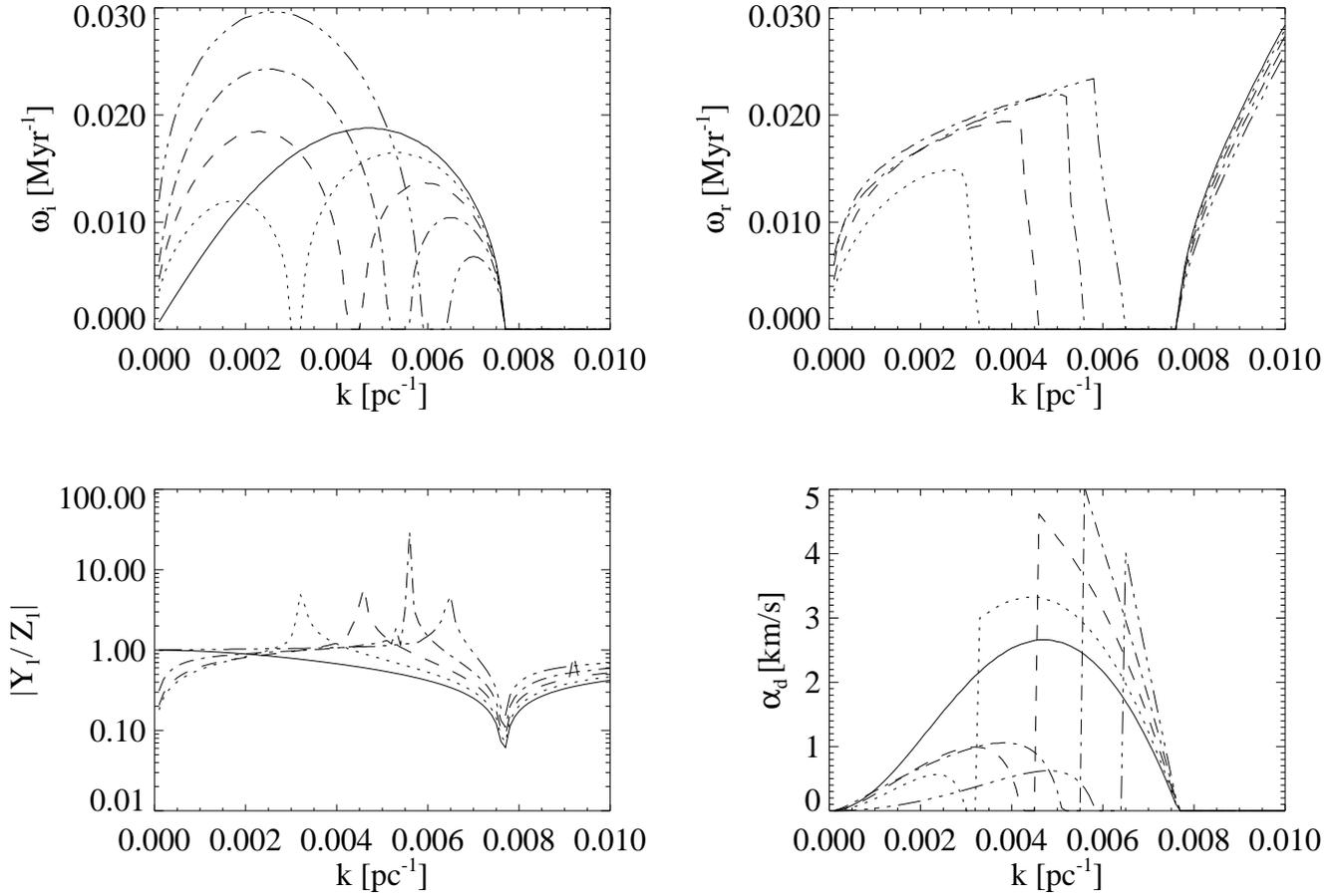}
\caption[]{
The dependence of growth rate, frequency, $|Y_1/Z_1|$ amplitude ratio and
$\alpha_d$ on wavenumber $k$ for fixed $\Omega = \Omega_G$ and the differential
force related to $A/\Omega=0$ (full line), -1/4 (dotted line) -1/2 (dashed
line), -3/4 (dot-dashed line) and -1 (3*dot-dashed line).
}
\end{figure*}
%%%%%%%%%%%%%%%%%%%%%%%% END OF FIG.8 %%%%%%%%%%%%%%%%%%%%%%%%%%%%%
The results can be commented as follows:
\begin{enumerate}
\item The curves of growth rate are split into 2 separate parts
in each case.
The minimum between them is placed at the wavenumbers $k = k_{crit}$ related to
the balance between:  the radial component of the magnetic tension, the radial
component of the Coriolis force and the differential force.  The right hand
parts are associated with vanishing oscillation frequencies and the left hand
parts with non-vanishing ones (see the right top panel).  This means that in
the limit of small wavenumbers we deal with propagating waves in contrast to
the stationary waves in the range between $k_{crit}$ and $k_{marg}$, where
$k_{marg}$ is assigned to the usual marginal stability point.  In this respect
the Parker--shearing instability is in the range $(k_{crit},k_{marg})$ similar
to the Parker instability without shear.  The typical phase speeds $\omega_r/k$
are close to the Alfven speed $v_A \sim 10 {\rm km/s}$.

It would be convenient to introduce names of the right-hand and left-hand
ranges.  We would suggest splitting the name {\em
Parker-shearing instability} into {\em the Parker range} for $k_{crit} < k <
k_{marg} $ and {\em the shearing range} for $ 0 < k < k_{crit} $.

\item The modulus of the ratio of radial to vertical displacement amplitudes
$|Y_1/Z_1|$, is strongly modified with respect to the case without shear.  The
reduction of the restoring
radial magnetic tension by the opposite differential force results in enhanced
radial (Y) excursions of the flux tube for perturbations placed near the
critical wavenumber. One should notice that only the imaginary part $\Im
(Y_1/Z_1)$ contributes to $\alpha_d$. The $\Im(Y_1/Z_1)$ is identical to
$|Y_1/Z_1|$ in the Parker range and is much smaller than $|Y_1/Z_1|$ in the
shearing range because of an additional phase shift between the radial and
vertical displacements. This gives rise to the apparent jump of $\alpha_d$ at
the critical wavenumber $k_{crit}$.
In the Parker range the $\alpha$-effect is strongly magnified and in the
shearing range it is strongly diminished with respect to the vanishing shear
case.
\item We can  expect also an enhanced
horizontal diffusion near $k_{crit}$ since
the radial displacements of the flux tube are approximately ten times larger
than the vertical ones. Then, the horizontal diffusivity can be even two orders
of magnitude larger than the vertical diffusivity.  This feature makes a big
qualitative difference with respect to the superbubble model by Ferriere
(1996), where the vertical diffusivity is comparable or larger to the
horizontal one depending on the galactic height.

\end{enumerate}

\subsection{Periodic variations of differential force due to the
linear density wave}

Let us try to imagine what happens in the presence of density waves.  Let
us assume that a single wavelength of the Parker--shearing instability
dominates, related to $k_0 = 0.005 {\rm pc}^{-1}$.  We expect that the density
wave introduces periodic modification of the differential force around the mean
value resulting from the axisymmetric rotation.  Assume that the amplitude of
these oscillations is sufficiently large to cause  the boundary between the
Parker range and the shearing range to oscillate around $k_0$ depending on the
phase of density wave.
{\bf Then, we can expect a switch--on and --off mechanism
of the $\alpha$-effect.}

Now, we shall recall some basic properties of density waves.
Stellar density waves in the galactic disc introduce periodic perturbations to
the axisymmetric gravitational potential $\Phi_0(R)$ (see eg.  Binney \&
Tremaine 1987, p.  386-388).  In the linear approximation and with the
assumption that the density waves are tightly wound, these perturbations have
the following form
\begin{equation}
\Phi_1(R,\varphi) = F \cos(k_{dw} R + m \varphi),
\end{equation}
where $k_{dw}$ is the radial wavenumber of the density wave, $m$ is the
azimuthal wavenumber of the density wave, $R$ and $\varphi$ are the radius and
the azimuthal angle in galactic disc.  The amplitude $F$ can be estimated from
the condition for cloud--cloud collisions (eqn.  (6-75) of Binney \& Tremaine
1987).  Taking into account $g = 1$ for which the collisions start to occur, we
can estimate

\begin{equation}
F \simeq \frac{\mid \kappa_0^2 - \omega_{dw}^2 \mid}{k_{dw}^2}.
\end{equation}
If we introduce a typical value of the epicyclic frequency $\kappa_0 = 1.2
\Omega_G$, and the oscillation frequency of the density wave
$\omega_{dw} = 0.75 \kappa_0$ and $k_{dw} = 2\cdot 10^{-3} {\rm pc}^{-1}$ then
\begin{equation}
F = 150 {\rm\frac{pc^2}{Myr^2}}
\end{equation}
The gravitational acceleration due to this potential is
\begin{equation}
a_{r1} = -\frac{\partial \Phi_1}{\partial R} = k_{dw} F \sin (k_{dw} R + m
\varphi)
\end{equation}
For our purposes it will be necessary to know the differential acceleration
due to the density wave acting on 2 points displaced by $\Delta R$ in the
radial direction
\begin{equation}
\Delta  a_{r1} = \frac{\partial a_{r1}}{\partial R} \Delta R
= k_{dw}^2 F \cos(k_{dw} R + m \varphi)  \Delta R
\end{equation}
The constant
\begin{equation}
A_{r1}\equiv k_{dw}^2 F
\end{equation}
describes the amplitude of the oscillations of the differential force.
Substituting the numerical values for $k_{dw}$ and $F$ we obtain
\begin{equation}
A_{r1} = 0.6 \cdot 10^{-3} Myr^{-2},
\end{equation}
while the differential acceleration resulting from axisymmetric rotation is
described by the analogous constant $A_{r0} = -4 A \Omega \simeq 2 \div 3\cdot
10^{-3}
Myr^{-2}$ depending on the type of the rotation curve.  We notice that the
'top--bottom' amplitude of oscillations of the differential force is of the
order of 50\% of the mean value coming from the axisymmetric
differential rotation.

{ It appears that the differential acceleration due to
the density waves is comparable to the differential acceleration due to the
axisymmetric differential rotation even if the perturbation of interstellar
gas is   at the limit of linear regime. This limit is
related to approximately 1\%
magnitude of perturbation of the axisymmetric gravitational potential (Roberts,
1969; Shu et al. 1973). From the above considerations one can easily derive the
relevant relation
\begin{equation}
\frac{A_{r1}}{A_{r0}} = \frac{\pi^2 R^2}{\lambda^2_{dw}}
\frac{\Omega}{-A} \frac{\Phi_1}{\Phi_0}
\end{equation}
where $\lambda_{dw}$ is the radial wavenumber of the density wave and
$\Phi_1/\Phi_0$ is the relative  magnitude of the axisymmetric gravitational
potential perturbation. This is obvious that typically the
density wave has the wavelength much shorter than the value of the radial
coordinate at a given point in the galactic disc, and then $A_{r1}/A_{r0}$ can
be be larger than $\Phi_1/\Phi_0$ even by a factor of a few tens.}

{\bf This implies that the switch--on and --off mechanism
of the $\alpha$-effect is really possible in observed galaxies.}

Let us try to figure out what is the phase difference between the oscillations
of differential force and the density maxima of spiral arms.  The density
maxima coincide with minima of the periodic component of gravitational
potential.  The periodic acceleration is directed toward the density
concentrations and the derivative of the periodic acceleration is in phase with
the potential perturbations.  This means that the density wave tends to
decrease the differential force in arms and increase it in the interarm
regions.  As we already suggested this effect can cause that a perturbation
with some dominating wavenumber $k_0$ moves from the Parker range to the
shearing range periodically.  Now, we have fixed that the Parker range
coincides with arms and the shearing range with interarm regions.
The validity of the current formulation of the problem depends on time available
between the passages of two neighboring arms of the galactic spiral structure.
One can say that our approach is valid if the growth time of the Parker
instability is much shorter than the density wave oscillation period.
This criterion is well fulfilled in the limit of weak magnetic field strength
and high cosmic ray pressure, but it is fulfilled only marginally if the
magnetic pressure is comparable to the gas pressure.

\section{Conclusions}

Conclusions which follow from our calculations can be splited in two groups.
The first group is related mainly to the cosmological generation of the
galactic magnetic fields due to the Parker instability. In this case the
presence of shear magnifies the dynamo effect, however it is of secondary
importance. The second group concerns the contemporary evolution of magnetic
field in galaxies. Some observed features of magnetic field in nearby galaxies
seem to be a signature of the active role of differential forces due to
the axisymmetric shear and the density waves as well.

The first group of conclusions is as follows:
\begin{enumerate}
\item Contrary to the common  opinion that the Parker instability was
ineffective in the past we note that weak magnetic field enables a fast growth
of Parker modes due to a lack of magnetic tension if only the cosmic ray
pressure was high enough.
\item For weak magnetic fields the  strong cosmic ray pressure cause the
main contribution to the dynamo $\alpha$-effect to come from waves  shorter
then
100 pc, which is good for the ``mean field dynamo theory'' since the shorter
waves mean a better statistics.
\item If the  magnetic field strength exceeds a few $\mu$G, the
time necessary for $\alpha_d$ to reach 1 km/s exceeds the galactic rotation
period and the discussed
process can only sustain the existent magnetic field strength.
\end{enumerate}

{\bf Based on these conclusions we can formulate a hypothesis that during
galaxy formation stars started to form in the presence of a very weak magnetic
field, producing an excess of cosmic ray pressure over the magnetic pressure.
The excess of cosmic rays forced a strong dynamo action converting the
exceeding cosmic ray energy to the magnetic energy.  The contemporary
equipartition is the result of this process.}

This scenario finds its convincing proof in the observational fact that
at high redshifts the number of blue objects, i.e. excessively starbursting galaxies,
is much larger than today (e.g. Lilly 1993). Just recently, an analysis of the
deepest available images of the sky, obtained by the Hubble Space Telescope
reveals a large number of candidate high redshift galaxies, up to redshifts $z\ge 6$.
The high-redshift objects are interpreted  as regions of intense star formation
associated with the progenitors of present-day normal galaxies, at epochs that may
reach back 95\% of the time to the Big Bang (Lanzetta et al. 1996).

We note that the effective action of a "Parker-dynamo" would also offer a
solution for the problem of magnetization of the intergalactic medium.
Especially the origin of magnetic fields in the radio halos of galaxy clusters,
like the Coma-cluster is still an open question (Burns et al.  1992).  Our
scenario may provide the backstage for magnetizing large cosmic volumes via
galactic winds.  If all galaxies go through a phase of intense star formation,
they will drive galactic winds, which like in the case of the standard
starburster M82 will have strong winds, which transport relativistic electrons
and magnetic fields into the intergalactic medium (Kronberg and Lesch 1996).
The radio halo of M82 is about 10 kpc in radius with a mean field of about
$10\mu$G (Reuter et al.  1992).

The second group of conclusions is related closely to the galactic dynamics.
{\bf We predict a strong $\alpha$ effect in arms and very weak one between
arms}.  Let us recollect that the magnified $\alpha$-effect in the Parker range
near the critical wavenumber $k_{crit}$ is the result of shearing forces which
counteract the radial magnetic tension and enable large amplitudes of radial
deformations of the azimuthal flux tube.  We expect also an enhanced amount of
the chaotic component of magnetic field in arms.

Then, we can propose a new galactic dynamo model relying on a
cyclic process composed of the action of galactic axisymmetric differential
rotation, density waves, Parker-shearing instability and
magnetic reconnection.  The cyclic process is as follows:  The $\alpha$-effect
due to the Parker-shearing instability is controlled by the magnitude of shear.
We postulate that the initial state of perturbation can be determined by small
irregularities in the interarm region (like e.g the velocities of small
molecular clouds).  Then the approaching arm induces a strong $\alpha$-effect
due to Parker-shearing instability and in consequence the dynamo action.  The
magnetic field becomes irregular due to the vertical and horizontal
displacements of flux tubes.  As it follows from Fig.~8 the horizontal
displacements exceed significantly the vertical ones.  When the arm passes
considered region by, the $\alpha$-effect becomes suppressed again.  Then the
enhanced shear of interarm region makes order in horizontal structure of
magnetic field and reconnection removes the vertical irregularities, thus our
model predicts a more regular magnetic field in the interarm regions.
We are in
the state which is similar to the initial one with the difference in the
strength of magnetic field which is stronger now due to the recent dynamo
action.  The characteristic feature of the proposed model is the dynamo action
(strictly speaking the $\alpha$-effect)  operating exclusively in spiral
arms.

These results perfectly explain observational results of Beck \& Hornes
(1996). The sharp peaks of the total power and the polarized
emission presented in their Fig. 3 suggest that the switch--on and --off
mechanism of the $\alpha$-effect of our model takes place in NGC 6946.

Moreover, the mentioned observations seem to contradict the critics of the
dynamo mechanism by Kulsrud and Anderson (1992). The fast uniformization of the nonuniform
magnetic field in spiral arms has the observed time scale order of the fraction
of galactic rotation period. This means that processes faster than the
ambipolar diffusion play a role in dissipation of irregular component of the
galactic magnetic field. Since our model represents a kind of the fast dynamo
(Parker, 1992) we postulate that the magnetic reconnection can efficiently
remove the short wavelength irregularities of flux tubes.

The application of two intrinsic instabilities of magnetized
differentially rotating galactic discs leads to interesting conclusions
about the "magnetic history" of the Universe. The magnetic fields in the high-redshift
objects is already as strong as the magnetic fields in nearby galaxies.
Thus,
energetically nothing has happened for many Gigayears in disc galaxies, their
magnetic energy density stays at the level of equipartition. This already indicates
that global saturation mechanisms like the Parker-instability are at work in these
objects. Adding observed properties of the high-redshift objects, namely
intense star formation, accompanied by effective production of cosmic rays,
we have developed a very promising scenario for the evolution of magnetic fields
in galaxies, which is able to reproduce the magnetic structure
in nearby galaxies and the saturation of magnetic energy density during galactic
evolution.

\begin{acknowledgements}
It is a pleasure to thank Thierry Foglizzo for the very detailed and fruitful
discussions during the work on this paper and for his critical reading of the
manuscript. We thank also Marek Urbanik and an anonymous referee for their
valuable comments.
MH thanks the Max-Planck-Institut f\"ur Astrophysik, Garching, Germany for a
kind hospitality.
This work was supported by the grant from Polish Committee for
Scientific Research (KBN), grant no. PB 0479/P3/94/07.
\end{acknowledgements}

\end{document}